\documentclass[preprint]{aastex}

\newcommand{\beq}{\begin{equation}}
\newcommand{\eeq}{\end{equation}}
\newcommand{\bea}{\begin{eqnarray}}
\newcommand{\eea}{\end{eqnarray}}
\newcommand{\nn}{\nonumber}
\newcommand{\um}{^{\mu}}
\newcommand{\un}{^{\nu}}
\newcommand{\dm}{_{\mu}}
\newcommand{\dn}{_{\nu}}
\newcommand{\dmn}{_{\mu\nu}}
\newcommand{\umn}{^{\mu\nu}}

\newcommand{\cF}{{\cal F}}
\newcommand{\cA}{{\cal A}}
\newcommand{\cJ}{{\cal J}}
\def\pa{\partial}
\def\bB{\mbox{\boldmath$B$}}
\def\bE{\mbox{\boldmath$E$}}
\def\bJ{\mbox{\boldmath$J$}}
\def\bA{\mbox{\boldmath$A$}}
\def\bP{\mbox{\boldmath$P$}}
\def\bQ{\mbox{\boldmath$Q$}}
\def\bV{\mbox{\boldmath$V$}}
\def\bbe{\mbox{\boldmath$\beta$}}
\def\rg{{r_g\over r}}
\def\rs{r_{\ast}}
\shorttitle{General Relativistic Particle Acceleration in Pulsar Polar Cap}
\shortauthors{Sakai \& Shibata}
\begin{document}

\title{General Relativistic Electromagnetism and Particle Acceleration 
in Pulsar Polar Cap}
\author{Nobuyuki Sakai}
\affil{Faculty of Education, Yamagata University, Yamagata 990-8560, Japan}
\email{sakai@ke-sci.kj.yamagata-u.ac.jp}
\and
\author{Shinpei Shibata}
\affil{Faculty of Science, Yamagata University, Yamagata 990-8560, Japan}
\email{shibata@sci.kj.yamagata-u.ac.jp}

\begin{abstract}
We reconstruct a 3+1 formalism of general relativistic 
electromagnetism, and derive the equations of motion of charged 
particles in the pulsar magnetosphere, taking account of the inclination 
between the rotation axis and the magnetic axis. Apart from the 
previous works where space charge is evaluated by assuming the flow 
velocity being the light speed, we analyze particle 
motion in the polar cap, finding that gravity changes significantly its 
dynamics and the condition for acceleration.
\end{abstract}


%%%%%%%%%%%%%%%%%%%%%%%%%%%%%%%%%%%%%%%
\section{Introduction}
%%%%%%%%%%%%%%%%%%%%%%%%%%%%%%%%%%%%%%%

The origin of radio emission from pulsars remains a great mystery. One of 
the likely scenarios is that particles are accelerated along open magnetic 
field lines and emit $\gamma$-rays which subsequently convert into 
electron-positron pairs under a strong magnetic field. A combination 
of the primary beam and pair plasma provide the radio emission 
mechanism \cite{Mel}.
The process of particle acceleration has been intensively studied in flat 
spacetime \cite{FAS,AS,Mes81,Shi91,Shi95,Shi97}.
The field-aligned electric field is driven by deviation of the space 
charge from Goldreich-Julian charge density \cite{GJ}, which is determined by 
the magnetic field geometry. Therefore, general relativistic (GR) effects
on the field geometry are crucial for the formation of the 
field-aligned electric field and particle acceleration.

Muslimov \& Tsygan (1992) initiated the GR analysis of electromagnetic 
fields around a pulsar; they solved Maxwell equations on the 
assumption that the particles move with the light speed. Shibata 
showed, however, that their assumption of constancy of particle speed was not 
always true \cite{Shi91,Shi97}: 
the particle motion and space charge show oscillatory behavior even 
on field lines curving toward the rotation axis if the assumed current 
density is sub-critical.
Mestel (1996) extended Shibata's equations of motion to include GR 
effects, though his analysis is not complete.
Recently Harding and Muslimov (2001) extended Muslimov-Tsygan's 
analysis to include the pair-creation dynamics.

In this paper we derive the basic equations more rigorously and generally:
we depart Muslimov-Tsygan (and Harding-Muslimov) model by solving equations 
of motion for the particles, and correct Mestel's equation and extend it 
to include the inclination angle between the magnetic axis and the 
rotation angle. Next we solve the electric field together with the 
particle motion for the region near the magnetic pole just above the surface. 
It turns out that the effect of gravity on particle dynamics cannot be 
ignored, and hence the condition for acceleration is also modified. 
As pointed by Arons (1997), local deficit of space charge takes place 
on most of the field lines. In particular, particle dynamics on the 
``away'' field lines from the rotation axis is changed qualitatively by 
gravity if the current density is less than the critical value. It is 
also true that oscillatory solution exists even with GR effect. We 
find that some oscillatory solutions disappear above a certain height. 

This paper is organized as follows. In section 2, we reconstruct a 3+1 
formalism of GR electromagnetism. In section 3, we derive 
equations of motion of charged particles around a magnetized 
rotating star. In section 4, solving the equations of motion, we analyze 
particle acceleration in the polar cap. Conclusions and discussions are given in section 5.
We use the unit $c=1$.

%%%%%%%%%%%%%%%%%%%%%%%%%%%%%%%%%%%%%%%
\section{3+1 Electromagnetic Equations: General Formalism}
%%%%%%%%%%%%%%%%%%%%%%%%%%%%%%%%%%%%%%%

A 3+1 split of Maxwell equations were made by Landau \& Lifshitz 
(1975) and Thorne \& Macdonald (1982). Landau \& Lifshitz derived 3+1 Maxwell equations 
in a straightforward manner, which we essentially follow. 
They introduced a metric
\footnote{In this paper Greek letters run from 0 to 3, and Latin letters run from 1 to 3, 
contrary to the convention of Landau \& Lifshitz (1975).},
\beq\label{LLds}
ds^2=g\dmn dx\um dx\un
=-h(dt-g_i dx^i)^2+\gamma_{ij}dx^idx^j,
\eeq\beq
h\equiv-g_{00},~~ g_i\equiv{g_{0i}/h},~~ \gamma^{ij}\equiv g^{ij},
\eeq
which is formally in a 3+1 form.
However, it is hard to understand how the entire 4D spacetime ($x\um,g\dmn$) is 
decomposed into 3D spatial hypersurfaces ($x^i,q_{ij}$) because each hypersurface 
does not correspond to $t=$constant.

On the other hand, Thorne \& Macdonald adopted the metric used in the ADM formalism:
\beq\label{ds}
ds^2=g\dmn dx\um dx\un=-\alpha^2dt^2+q_{ij}(dx^i+\beta^idt)(dx^j+\beta^jdt),
\eeq\beq
\alpha^2\equiv-{1/g_{00}},~~ \beta_i\equiv g_{0i},~~ 
q_{ij}\equiv g_{ij}, 
\eeq
where $\alpha$ and $\beta$ are called the lapse function and the shift vector, 
respectively. With this metric one can understand the concept of the 3+1 
decomposition easily, as we describe below. However, they adopted the ``congruence'' 
approach and wrote down the equation with 
coordinate-free language such as expansion and shear of the fiducial 
observers. Although their equations are powerful for some theoretical arguments, 
one cannot avoid using coordinates to solve them in most cases.

For the purpose of our study, we combine and modify the two methods: we adopt the 
ADM formalism to decompose a spacetime and electromagnetic fields into 
a 3+1 form, and write down 3+1 electromagnetic equations in a straightforward manner.

First, with the metric (\ref{ds}), we introduce the fiducial observers with the 4-velocity,
\beq\label{u}
u\dm=(-\alpha,0,0,0), 
\eeq
and the spacelike hypersurface $\Sigma$ which is orthogonal to $u\um$ 
at each time $t$. Because $\Sigma(t)$ is characterized by a $t=$ constant 
hypersurface, $x^i$ and $q_{ij}$ are 
the 3-coordinates and the 3-metric of $\Sigma(t)$, respectively. 
We define the projection tensor, which projects any 4-vector or tensor 
into $\Sigma(t)$, as
\beq
h\umn\equiv g\umn+u\um u\un.
\eeq

Electromagnetic quantities are originally in the 3+1 form, i.e., 
3-vectors and scalars: electric field $\bE$, magnetic field $\bB$, 
charge density $\rho$, current density $\bJ$, scalar potential $\phi$, 
and vector potential $\bA$. We define those quantities on $\Sigma(t)$, 
i.e., those quantities are supposed to be measured by the 
observers with the velocity $u\um$.
Note that the ``ordinary'' components of a vector $\bP$ are neither the covariant 
components $P_i$ nor the contravariant components $P^i$ \cite{Wei}.
If the 3-coordinates $x^i$ are orthogonal, i.e., 
$q_{ij}=h_{(i)}\delta_{ij}$, the ordinary components $P_{(i)}$ are
\beq
P_{(i)}=h_{(i)}P^i={P_i/h_{(i)}}.
\eeq
As long as we adopt the 4-velocity (\ref{u}), which specifies 
$\Sigma$, we can convert any 3-vector $\bP$ on $\Sigma$ is converted into 
a 4-vector $P\um$ in a natural way,
\beq
P\um=(0,~P^i).
\eeq
It is easy to confirm that $P\um$ is the vector on $\Sigma$: $P\um u\dm=0$ and $P\um h\dm^i=P^i$.

Let us reconstruct the electromagnetic field $\cF\umn$, the charge-current 
vector ${\cJ\um}$ and the 4-potential $\cA\um$ from the quantities defined 
on $\Sigma$ and $u\um$ \cite{TM}:
\bea\label{defF}
&&\cF\umn=u\um E\un-u\un E\um
+\epsilon^{\mu\nu\lambda\sigma}u_{\lambda}B_{\sigma}, \\
&&\cJ\um=\rho u\um+J\um, ~~
\cA\um=\phi u\um+A\um,
\label{defJA}\eea
where $\epsilon^{\mu\nu\lambda\sigma}$ is the Levi-Civita tensor with
$\epsilon^{0123}=1/\sqrt{-g}$. We can invert these relations as
\bea
E^i=\cF^{i\nu} u\dn, &~~& 
B^i=-\frac12\epsilon^{i\nu\lambda\sigma}u\dn \cF_{\lambda\sigma}, \\
\rho=-\cJ\um u\dm, &~~& J^i=h^i\dn\cJ\un, \\
\phi=-\cA\um u\dm, &~~& A^i=h^i\dn \cA\un.
\eea

The 4D expressions of Maxwell equations and the continuity equation 
are
\beq\label{4DMax}
\cF\umn_{~~;\nu}=4\pi\cJ\um,~~ \cF_{[\mu\nu;\lambda]}=0,~~
\cJ\um_{~;\mu}=0.
\eeq
Substituting (\ref{defF}) and (\ref{defJA}) into (\ref{4DMax}), we 
obtain 3+1 Maxwell equations:
\bea\label{divB}&&
{\rm div}\bB=0,
\\\label{rotE}&&
{1\over\sqrt{q}}\pa_t(\sqrt{q}B^i)
+[{\rm rot}(\alpha\bE+\bbe\times\bB)]^i=0, 
\\\label{divE}&&
{\rm div}\bE=4\pi\rho,
\\\label{rotB}&&
-{1\over\sqrt{q}}\pa_t(\sqrt{q}E^i)
+[{\rm rot}(\alpha\bB+\bbe\times\bE)]^i=4\pi\alpha J^i,\label{M34}
\\\label{CC}&&
{1\over\sqrt{q}}\pa_t(\sqrt{q}\rho)+{\rm div}(\alpha\bJ)=0,
\eea
where cross product, divergence and rotation are defined as
\beq
[\bQ\times\bP]^i \equiv \epsilon^{ijk} Q_j P_k, ~~
{\rm div}\bP \equiv{1\over\sqrt{q}}\pa_i(\sqrt{q}P^i), ~~
[{\rm rot}\bP]^i\equiv\epsilon^{ijk}\pa_jP_k,
\eeq
and $\epsilon^{ijk}$ is the spatial Levi-Civita tensor with 
$\epsilon^{123}=1/\sqrt{q}$.

To complete 3+1 electromagnetic equations, we rewrite 
the equation of motion of a charged particle. We denotes its mass 
by $m$, its charge by $e$, its proper time by $\tau$ and its 4-velocity by 
$v\um\equiv dx\um/d\tau$. Then the equation of motion is written as
\beq\label{4eom}
m\left({dv\um\over d\tau}+\Gamma\um_{\nu\lambda}v\un v^{\lambda}\right)
=e\cF\umn v\dn,
\eeq
where $\Gamma$ is the affine connection.
We define a scalar $\gamma$, the 3-velocity vector $V^i$ and its 
norm $V$ as
\beq\label{defV}
\gamma\equiv-v\um u\dm=\alpha v^0,~~
V^i\equiv{v\um h^i\dm\over\gamma}={v^i+\beta^iv^0\over\gamma},~~
V^2\equiv q_{ij}V^iV^j.
\eeq
From the normalization $v\dm v\um=-1$ and (\ref{defV}), we find
$\gamma={1/\sqrt{1-V^2}}$,
which implies that the scalar $\gamma$ is nothing but the Lorentz factor.
With these quantities and the 3+1 electromagnetic quantities we rewrite 
the the equation of motion (\ref{4eom}) as
\beq\label{eom1}
{d\over d\tau}\left\{\gamma(-\alpha+\bbe\cdot\bV)\right\}
-{\gamma^2\over\alpha}\left(-\pa_0\alpha+{\alpha\over2}V^iV^j\pa_0 q_{ij}
+\bV\cdot\pa_0\bbe\right)
={e\gamma\over m\alpha}\bE\cdot\bV,
\eeq\beq
{d\over d\tau}(\gamma V_i)
-{\gamma^2\over\alpha}\left(-\pa_i\alpha+{\alpha\over2}V^jV^k\pa_i q_{jk}
+\bV\cdot\pa_i\bbe\right)
={e\gamma\over m\alpha}\left\{\alpha\bE+\bV\times\bB-(\bE\cdot\bV)\bbe\right\}_i. 
\label{eom2}\eeq

%%%%%%%%%%%%%%%%%%%%%%%%%%%%%%%%%%%%%%%
\section {Electromagnetic Equations in Pulsar Magnetosphere}
%%%%%%%%%%%%%%%%%%%%%%%%%%%%%%%%%%%%%%%

We assume that the spacetime outside a neutron star is stationary and axisymmetric.
Because gravity of charged particles is negligible, the outer spacetime is 
described by Kerr metric. We adopt the spherical coordinates
$(\hat t,~\hat r,~\hat\theta,~\hat\varphi)$ so that $\hat\theta=0$ accords 
with the rotation axis. In the slow-rotation where we take up to the first order
of the angular momentum of the star $L$, the spacetime is described by the metric,
\beq\label{ds2}
\hat\alpha^2=F(\hat r)\equiv1-{\hat r_g\over\hat r},~~
\hat\beta^{\hat\varphi}=-\omega(\hat r)\equiv-{2L\over\hat r^3},~~
\hat q_{ij}={\rm diag}(F^{-1}(\hat r),~\hat r^2,~\hat r^2\sin^2\hat\theta),
\eeq
where $r_g$ is the gravitational radius.

We consider the two-step coordinate transformations. First, we 
introduce the ``corotating coordinates'' as 
\beq
\hat\varphi\longrightarrow\hat\varphi-\Omega\hat t,
\eeq
where $\Omega$ is the angular velocity of the star \cite{LL}. The
shift vector is, accordingly, transformed as
\beq
\hat\beta^{\hat\varphi}\longrightarrow\hat\beta^{\hat\varphi}=\Omega-\omega.
\eeq
Second, we move on to the ``magnetic coordinates'' $(t,~r,~\theta,~\varphi)$, 
where the magnetic dipole axis accords with $\theta=0$.
Supposing the magnetic axis is at $\hat\theta=\chi,~\hat\varphi=0$, the 
coordinate transformation from the corotating coordinate to the 
(corotating) magnetic coordinate is given by
\bea
&&\hat x=\hat r\sin\hat\theta\cos\hat\varphi,~~
\hat y=\hat r\sin\hat\theta\sin\hat\varphi,~~
\hat z=\hat r\cos\hat\theta,~~
\nn\\
&&x=\hat x\cos\chi+\hat z\sin\chi,~~
y=\hat y,~~
z=-\hat x\sin\chi+\hat z\cos\chi,
\\
&&x=r\sin\theta\cos\varphi,~~
y=r\sin\theta\sin\varphi,~~
z=r\cos\theta.
\nn\eea
The shift vector in (\ref{ds2}) is, accordingly, transformed into
\beq\label{ds31}
\beta^r=0,~~ \beta^{\theta}=-\sin\chi\sin\varphi(\Omega-\omega),~~
\beta^{\varphi}=(\cos\chi-\cot\theta\cos\varphi\sin\chi)(\Omega-\omega),
\eeq
while the lapse function and the spatial metric remain unchanged:
\beq\label{ds32}
\alpha^2(r)=F(r),~~ q_{ij}={\rm diag}(F^{-1}(r),~r^2,~r^2\sin\hat\theta).
\eeq

In this (and any) axisymmetric spacetime the observers with the velocity (\ref{u}) 
is called ``zero-angular-momentum observers'' (ZAMOs) because of 
vanishing their angular momentum per unit mass: $u\dm({\pa/\pa\varphi})\um=0$
\cite{TM}.

Now we analyze the basic equations (\ref{divB})-(\ref{CC}), (\ref{eom1}) 
and (\ref{eom2}) with the metric (\ref{ds31}) and (\ref{ds32}). First,
we suppose that the magnetic field distortion due to the external 
currents is negligibly small. Then the magnetic field is 
governed by (\ref{divB}) and (\ref{rotB}) with $\bJ=0$:
\beq\label{Bsol}
{\rm div}\bB=0,~~ {\rm rot}(\alpha\bB)=0.
\eeq
We adopt the dipole-like solution of (\ref{Bsol}) \cite{GO}:
\bea\label{Br}
B_{(r)}&\equiv&{B^r\over\sqrt{F}}
=B_0{f(r)\over f(\rs)}\left({\rs\over r}\right)^3\cos\theta
\approx B_0\left[1+\frac34\left(\rg-{r_g\over\rs}\right)\right]
\left({\rs\over r}\right)^3\cos\theta,
\\
B_{(\theta)}&\equiv&r^2B^{\theta}={B_0\over2}{\sqrt{F}\over f(\rs)}
\left[-2f(r)+3F^{-1}(r)\right]\left({\rs\over r}\right)^3\sin\theta
\nn\\
&\approx&B_0\left(1+{r_g\over2r}-{3r_g\over4\rs}\right)
\left({\rs\over r}\right)^3\sin\theta, \label{Bth}
\\
f(r)&\equiv&-3\left({r\over r_g}\right)^3
\left[\ln F(r)+\rg\left(1+{r_g\over2r}\right)\right]
\approx 1+{3r_g\over 4r},
\eea
where $r=\rs$ denotes the star surface, and the 
approximate expressions are obtained by expanding in powers of $r_g/r$
or $r_g/\rs$ \cite{KK}.

Equation (\ref{rotE}) is satisfied if there exists a potential $\Phi$ such as
\beq\label{Phi}
\alpha\bE+\bbe\times\bB=-\nabla\Phi.
\eeq
Substituting (\ref{Phi}) into the other of Maxwell equations (\ref{divE}), we obtain
\beq\label{divEax}
-{\rm div}\left({\nabla\Phi\over\alpha}\right)=4\pi(\rho-\rho_{GJ}),~~
4\pi\rho_{GJ}\equiv-{\rm div}\left({\bbe\over\alpha}\times\bB\right),
\eeq
where $\rho_{GJ}$ is the Goldreich-Julian density.
Each term is written explicitly as
\bea
{\rm div}\left({\nabla\Phi\over\alpha}\right)&=&
{\sqrt{F}\over r^2}{\pa\over\pa r}
\left(r^2{\pa\Phi\over\pa r}\right)+{1\over\sqrt{F}r^2}
\left[{1\over\sin\theta}{\pa\over\pa\theta}
         \left(\sin\theta{\pa\Phi\over\pa\theta}\right)
       +{1\over\sin^2\theta}{\pa^2\Phi\over\pa\varphi^2}\right],
\\
4\pi\rho_{GJ}&=&-\alpha\bB\cdot{\rm rot}{\bbe\over\alpha^2} \nn\\
&=&-{B_{(r)}\over\sqrt{F}\sin\theta}
\left\{ {\pa(\sin^2\theta\beta^{\varphi})\over\pa\theta}
+{\pa\beta^{\theta}\over\pa\varphi} \right\}
+B_{(\theta)}{F\sin\theta\over r}{\pa\over\pa r}
\left({r^2\beta^{\varphi}\over F}\right).
\eea

Let us consider the flow of charged particles. In the case under 
consideration, magnetic field dominates over electric field, where
particles are at the lowest Landau level and the inertial drift 
motion across the magnetic field is negligible. We may, therefore, 
regard that particles go along with magnetic field lines:
\beq\label{Vj}
\bV=\kappa\bB,~~ \bJ=\rho\bV=\rho\kappa\bB,
\eeq
where $\kappa=V/B\equiv|\bV|/|\bB|$ is a scalar function. 

Then continuity equation (\ref{CC}) implies \cite{MT,Mes96}
\beq
0={\rm div}(\alpha\bJ)={\rm 
div}(\alpha\rho\kappa\bB)=\bB\cdot\nabla(\alpha\rho\kappa),
\eeq
which reads
\beq
\alpha\rho\kappa={\sqrt{F}J\over B}={\rm constant~on~magnetic~field~
lines.}
\eeq

Equations (\ref{eom1}) and (\ref{eom2}) give the motion of 
charged particles. 
If we assume $B_{\varphi}=E_{\varphi}=0$, equation (\ref{Phi}) and 
(\ref{Vj}) give $V_{\varphi}=\pa_{\varphi}\Phi=0$. Using the 
relation
\beq\label{EV}
\bE\cdot\bV=-{\nabla\Phi\over\alpha}\cdot\bV
=-{1\over\alpha}V^i\pa_i\Phi
=-{1\over\alpha}\left({1\over\Gamma}{dx^i\over d\tau}+{\beta^i\over\alpha}
\right){\pa\Phi\over\pa x^i}
=-{1\over\sqrt{F}\Gamma}{d\Phi\over d\tau},
\eeq
we rewrite one of the equations of motion (\ref{eom1}) to
\beq\label{eom}
{d\over d\tau}(\sqrt{F}\Gamma)
=\frac em{1\over F}{d\Phi\over d\tau}. 
\eeq
The other equation (\ref{eom2}) need not be solved because the 
spatial trajectory is determined by (\ref{Vj}).

Finally, let us consider boundary conditions. We assume that the star crust 
is a perfect conductor, and hence particles on the surface do not suffer 
Lorentz force. Equations (\ref{eom2}), (\ref{Phi}), (\ref{Vj}) and 
(\ref{EV}) reduces this ideal-MHD condition to $\nabla\Phi=0$.
We also assume that there exist closed magnetic lines, where the ideal-MHD 
condition holds. 
Thus open field lines in the polar region have another boundary, 
defined by ``last open field lines'', $\theta=\theta_c(r)$. 
Therefore, our boundary conditions are $\Phi=\nabla\Phi=0$ on 
$r=r_{\ast}$ and on $\theta=\theta_c(r)$.

%%%%%%%%%%%%%%%%%%%%%%%%%%%%%%%%%%%%%%%
\section{Particle Acceleration in the Polar Cap}
%%%%%%%%%%%%%%%%%%%%%%%%%%%%%%%%%%%%%%%

We are interested in the ``polar cap'' region, $\theta<\theta_c(r)$, where particle 
acceleration may occur. For simplicity, we restrict ourselves to the region near 
the magnetic pole just above the surface, where the following approximations 
hold: (i) $\theta_c(r)\approx$ constant, (ii) $\theta\ll1$ (we take 
its first order), and (iii) $d/dr$ along the particle trajectory 
$\approx\pa/\pa r$. Therefore, we can apply our equation below only to
the region elongated less than the polar cap radius: $r-r_*<r\theta_c$.

As for the non-radial direction, we expand $\Phi$ as
\beq\label{Ylm}
\Phi=\sum_{l,m}\bar\Phi(r)Y_{lm}(\theta,\varphi).
\eeq
We only take the mode of the polar cap scale, $l\approx\pi/\theta_c$.

The above assumptions simplify (\ref{divEax}) as
\bea\label{divEpc}
&&-{\sqrt{F}\over r^2}{d\over dr}\left(r^2{d\Phi\over dr}\right)
+{l(l+1)\over\sqrt{F}r^2}\Phi
=4\pi\left({J\over V}-\rho_{GJ}\right), \\
&&4\pi\rho_{GJ}={2B\Omega\over\sqrt{F}}
\left\{\left(1-{\omega\over\Omega}\right)\cos\chi+\theta Q\cos\varphi\sin\chi\right\},
\label{GJ}\\
&&Q\equiv{r_g\over 2r}-{3\omega\over 2\Omega}+{r_g\omega\over r\Omega}
+{3\over 2fF}\left(1-{3r_g\over 2r}+{\omega\over 2\Omega}\right)
\approx\frac32-{11r_g\over r}-{3\omega\over 4\Omega},
\label{Q}\eea
where $B\equiv|\bB|=B_{(r)}+O(\theta^2)$
\footnote{
Equation (\ref{Q}) is slightly different from the expression obtained by 
Muslimov and Tsygan (1992) and Harding and Muslimov (2001):
\beq
Q={3r_g\over 2r}-{3\omega\over 2\Omega}
+{3\over 2fF}\left(1-{3r_g\over 2r}+{\omega\over 2\Omega}\right).
\eeq
We believe that there is an error in their calculation. 
Nevertheless this difference does not change results practically; 
because we have assumed $\theta\ll1$, all terms but 2/3 (the dominant 
one) in $Q$ are negligible in equation (\ref{GJ}).
}.
Equation (\ref{eom}) describes the 
time-evolution of particles; however, if particles make a stationary flow, we can read their spatial 
configuration by replacing the time $\tau$ with a spatial valuable $r$:
\beq\label{eompc}
{d\over dr}(\sqrt{F}\gamma)=\frac em{1\over F}{d\Phi\over dr}.
\eeq

Using the normalized variables \cite{Shi97},
\bea
j&\equiv&-{2\pi\sqrt{F(s)}J(s)\over\Omega B(s)}={\rm constant}, ~~
\phi(s)\equiv\frac em\Phi(s), ~~
s\equiv\sqrt{{2\Omega B_0 e\over mc^2}}r,\\
\bar j(s)&\equiv&-{2\pi\sqrt{F(s)}\rho_{GJ}(s)\over\Omega(s)B(s)}
=\left(1-{\omega(s)\over\Omega}\right)\cos\chi+\theta(s)Q(s)\cos\varphi\sin\chi,
\eea
we rewrite (\ref{divEpc}) and (\ref{eompc}) as
\bea\label{eq1}
{\sqrt{F}\over s^2}{d\over ds}\left(s^2{d\phi\over ds}\right)
-{l(l+1)\over\sqrt{F}s^2}\phi
&=&{B\over B_0\sqrt{F}}\left({j\over V}-\bar j\right), 
\\\label{eq2}
{d\over ds}(\sqrt{F}\gamma)&=&{1\over F}{d\phi\over ds}.
\eea
The particle trajectory is given by integrating 
$d\theta/dr=B^r/B^{\theta}$, resulting in
\beq
\theta(s)\approx\theta_{\ast}\sqrt{{s\over s_{\ast}}} ~~ {\rm for} ~~
\theta\ll1,
\eeq
where $\ast$ denotes a value at the star surface $r=r_{\ast}$ for any 
variable.
In the following analysis we fix some of the parameters: $r_{\ast}=10$km, 
$r_g=2$km, $\Omega=2\pi/0.1{\rm s},~ \omega_{\ast}/\Omega=0.1,~
\theta_c=\sqrt{\Omega r_{\ast}/c}\approx 0.046~(\pi/\theta_c\approx=69),~
\chi=30^{\circ}$ and $\gamma_{\ast}=1.00001$.

We can argue equation (\ref{eq1}) on the analogy of particle motion by regarding 
$\phi$ as position and $s$ as time.
At a given point (or time) $j>\bar j(s)$, the ``force term'' (LHS) of (\ref{eq1}) always 
positive; $\phi$ and $\gamma$ increase acceleratingly. If $j<\bar j(s)$, on the other hand, the 
force term takes both positive and negative, depending on $V$, and hence oscillating 
behavior is expected. If we set $r_g=\omega=0$ (no gravity) and $\chi=0$ (no inclination), 
we reproduce Shibata's result, $\bar j=1$ \cite{Shi97}. The expression 
of $\bar j$ in (\ref{eq1}) indicates that both effects of gravity and 
of inclination reduce $\bar j$, or equivalently, the critical value of $j$.

Numerical solutions of (\ref{eq1}) and (\ref{eq2}) in Figs.\ 1-3
verify the above arguments. It is apparent that gravity reduces the critical 
value of $j$:  monotonic increase of $\gamma$ appears with lower values of $j$ 
that in non-GR case.

Let us consider the dynamics for $j\approx\bar j_{\ast}$ carefully. To 
see the behavior analytically, we expand $\bar j$ with
$\xi\equiv s/s_{\ast}-1$ up to the first order, resulting in
\beq
\bar j(\xi)=\bar j_{\ast}\left[1+C\xi\right],
~~ C\approx 
3\left({\omega_{\ast}\over\Omega}+{\theta_{\ast}\over4}\cos\varphi\tan\chi\right).
\eeq
This implies that $\bar j$ increases or decreases according to the sign of $C$. 
For the particles on the magnetic axis ($\theta=0$), $\bar j$ is an increasing 
function due to the term $\omega_{\ast}/\Omega$ in GR case while it 
is constant in non-GR case.
This gravitational effect is illustrated in Fig.\ 1:
contrary to non-GR case (a), in GR case (b) oscillation for 
$j<\bar j_{\ast}$ does not continue to infinity 
but the particle velocity becomes $V=0$ at some point. Although the dynamics 
is independent of $\gamma_{\ast}$ in non-GR case, the 
ending point $V=0 ~(\gamma=1)$ depends on $\gamma_*$ in GR case: it becomes further for 
larger $\gamma_{\ast}$.

Dynamics of particles off the magnetic axis ($\theta\ne 0$) is more 
interesting. In the absence of gravity ($\omega_{\ast}=0$), the sign 
of $C$ depends simply on whether particles move ``toward'' to 
($\cos\varphi>0$) or ``away'' from ($\cos\varphi<0$) the rotation axis, 
as claimed by Shibata (1997). He showed, for example, that 
particles are accelerated after oscillation along ``away'' field 
lines if $j$ is slightly less than 1. This behavior is reproduced in 
Fig.\ 2(a). Gravity changes this behavior drastically. If we take 
$\omega_{\ast}/\Omega=0.1$ typically, $C$ becomes positive even on ``away'' 
field lines unless $\theta_{\ast}>0.4|\sec\varphi|\cot\chi>0.4\cot\chi$. 
Unless $\chi$ is large enough, in most small-$\theta$ region, $\bar j$ 
simply increases and acceleration after oscillation 
cannot occur. This argument is confirmed by the numerical result in 
Fig.\ 2(b).

For the particles moving ``toward'' the rotation axis, $\cos\varphi$ is 
positive, and therefore $\bar j$ always increases along the field lines. 
The results in Fig.\ 3 are also understandable. Among the solutions 
in Fig.\ 3(b), the solutions with $j=1.003\bar j_{\ast}$ could be 
realistic because particle energy becomes so large that 
electron-positron pairs are created with finite electric field.

%%%%%%%%%%%%%%%%%%%%%%%%%%%%%%%%%%%%%%%
\section {Conclusions and Discussions}
%%%%%%%%%%%%%%%%%%%%%%%%%%%%%%%%%%%%%%%

We have reconstructed a 3+1 formalism of general relativistic
electromagnetism, and derived the equations of motion of charged 
particles in pulsar magnetosphere. Our basic equations are the 
correct and generalized version of those of Muslimov \& Tsygan (1992) and 
of Mestel (1996) in the sense that the equations include arbitrary current
density, arbitrary velocity which satisfies the equation of motion, 
and arbitrary inclination angle between the rotation axis and the magnetic axis. 

We have solved our equations of motion together with the electric 
field structure along the magnetic field for the region near the magnetic 
pole just above the surface in our approximate method. We have found 
that the effect of gravity on particle dynamics cannot be ignored and hence the condition for 
acceleration is also modified. In particular, particle dynamics on the ``away'' field lines 
is changed qualitatively by gravity if the current density is sub-critical.

Monotonic increase of $\gamma$ with the boundary condition
$\bE\cdot\bB=0$ at the stellar surface and at infinity
is found with GR effect, and the achieved Lorentz factor
is significantly larger than the case of non-GR (Muslimov \&
Tsygan 1992). With sub-critical current densities, we find
oscillatory solutions even with GR effect. Since we take the
current densities as a free parameter, the local accelerator
model can be linked with the global model by adjusting the 
current density.

We have also found that some oscillatory solutions disappear above 
a certain height. This may indicate an intermittent flow of
charged particles rather than a steady outflow if the imposed
current density is smaller than a threshold. This may cause
pulse nulling.

As pointed by Arons (1997), an important GR effect is local increase of $\bar{j}$, 
which appears on most of field lines, no matter whether large scale
field curvature of magnetic field lines is "away" or "toward".
He suggested that this fact can be responsible for the axisymmetric
radio emission about the magnetic axis. If the current density
is taken as a free parameter, the super-critical current density
in the polar annuli can also be responsible for the distribution 
of radio emitting region.
We shall extend our analysis to a more global region to construct a pulsar model in 
the subsequent work.

\acknowledgements
N.S.\ thanks Yasufumi Kojima for discussions, and also Yukawa 
Institute for Theoretical Physics, where a part of this work has been 
done as a visitor program. 
This work was partially supported by the Grant-in-aid for Scientific 
Research Fund of Monbukagakusho (Nos.\ 13740139 and 12640229).

\newpage
%%%%%%%%%%%%%%%%%%%%%%%%%%%%%%%%%%%%%%%%%%%%%%%%%%%%%%%%%%%%%%%%%%%%%%%%

\clearpage

\begin{figure}
\figurenum{1}
\epsscale{.9}
\plotone{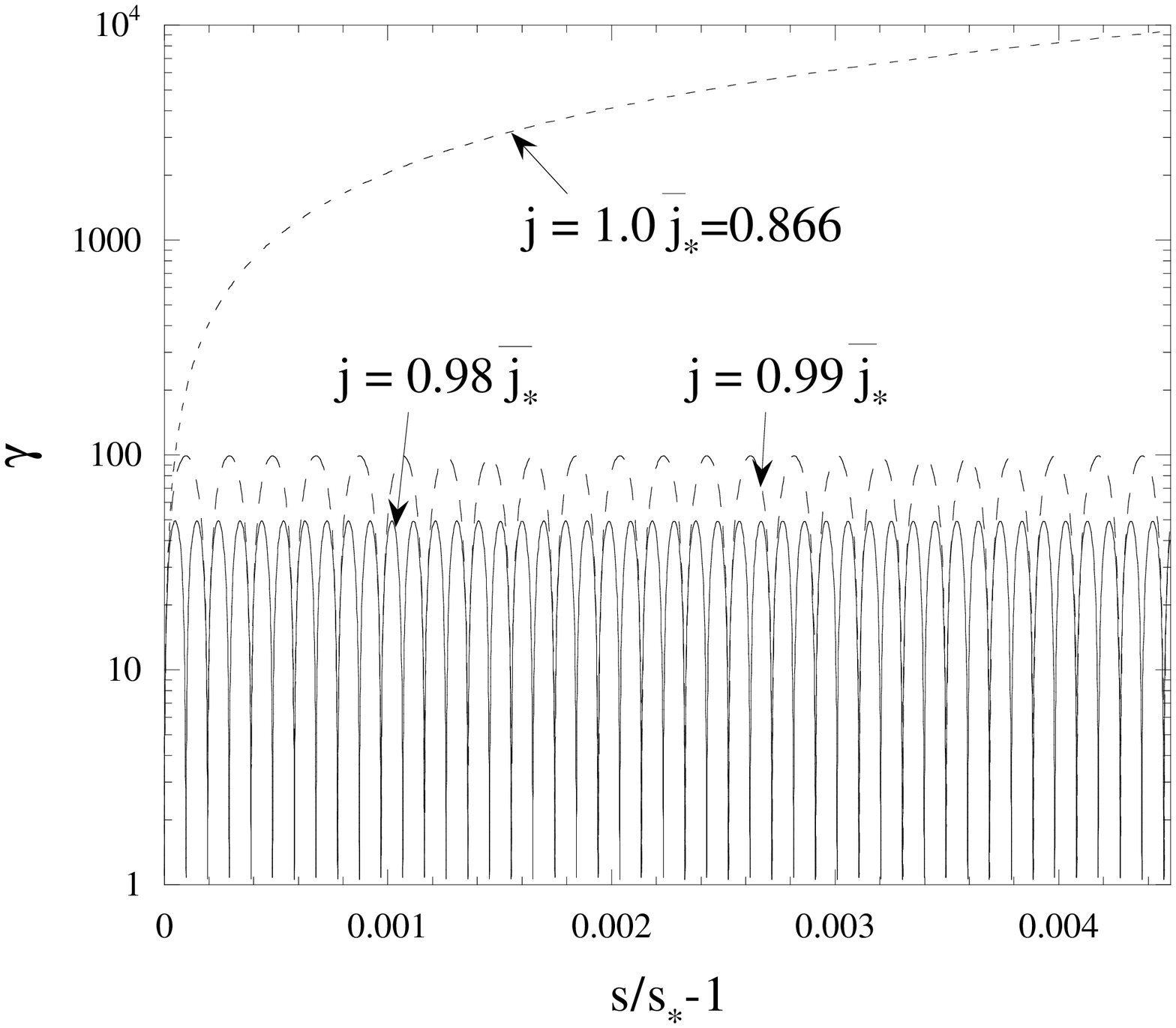}
\epsscale{.9}
\plotone{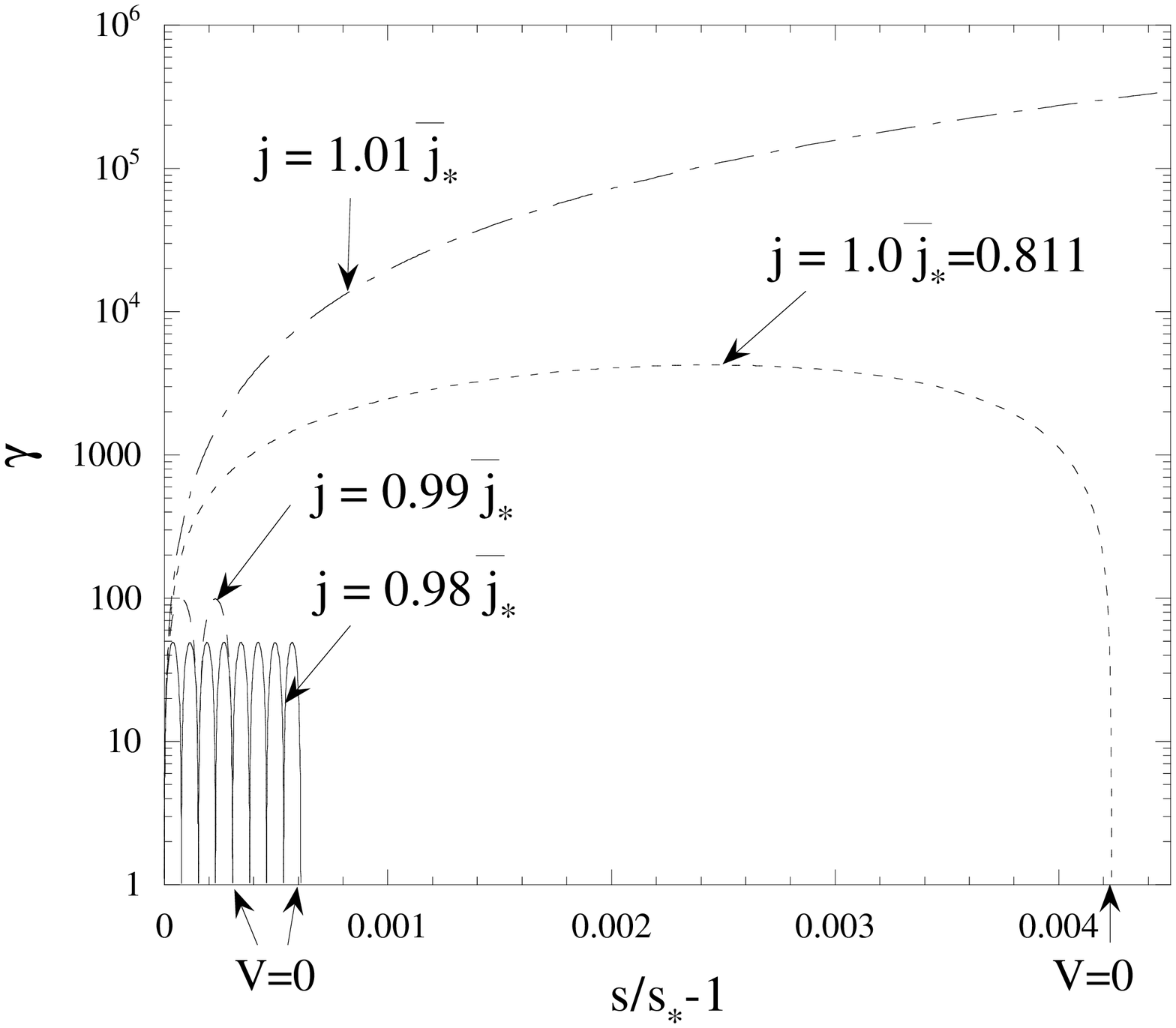}
\caption{Spatial distribution of $\gamma$ along magnetic field lines. We choose
$\theta=0$. (a) represents solutions for the flat spacetime with $r_g=\omega=0$.
$\gamma$ oscillates or increases monotonically according to $j$. (b) represents GR
solutions. In the case of $j<j_{\ast}$ oscillation 
eventually stops with $V$ approaching zero, in contrast with the non-GR case.}
\end{figure}

\begin{figure}
\figurenum{2}
\epsscale{.9}
\plotone{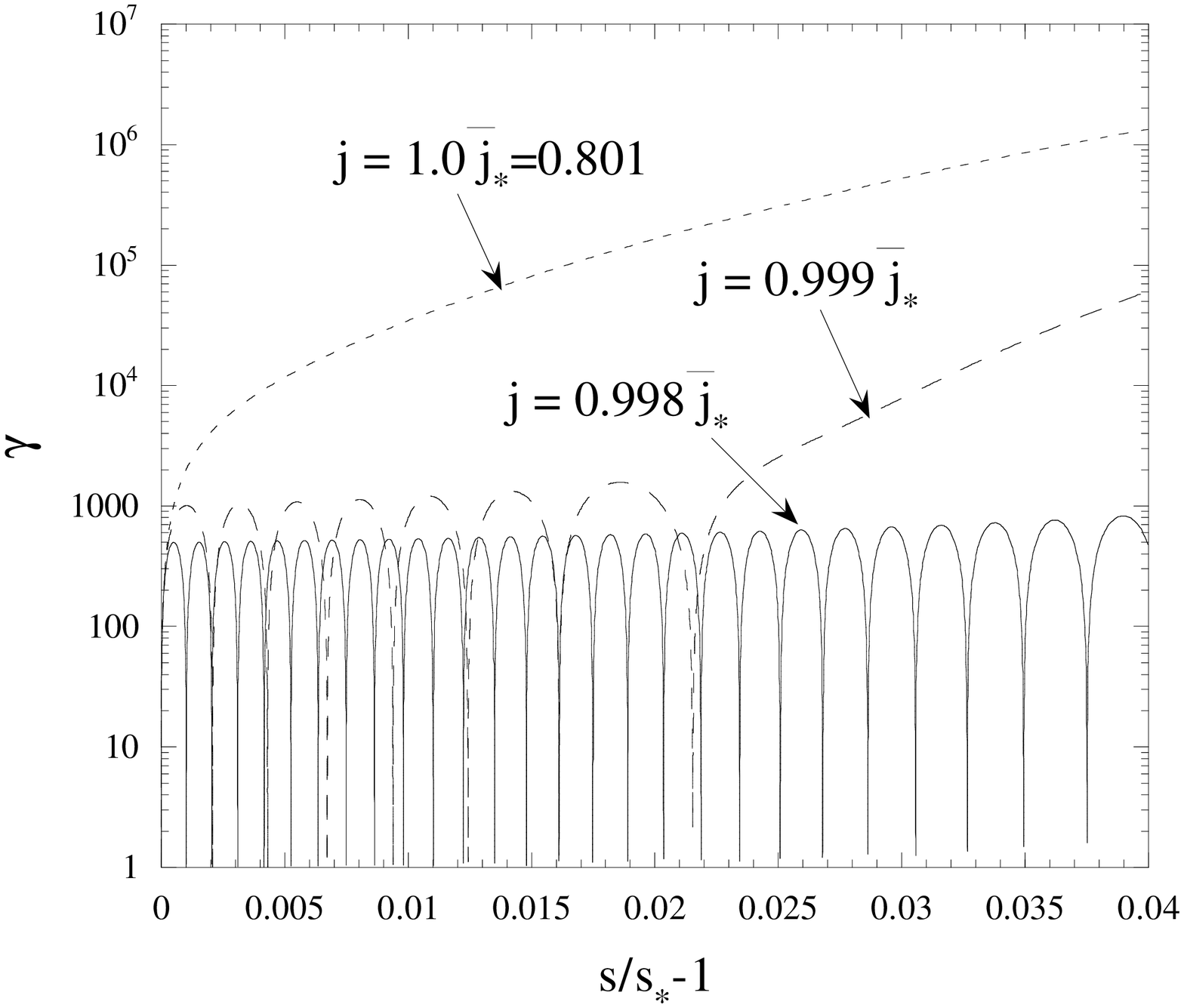}
\epsscale{.9}
\plotone{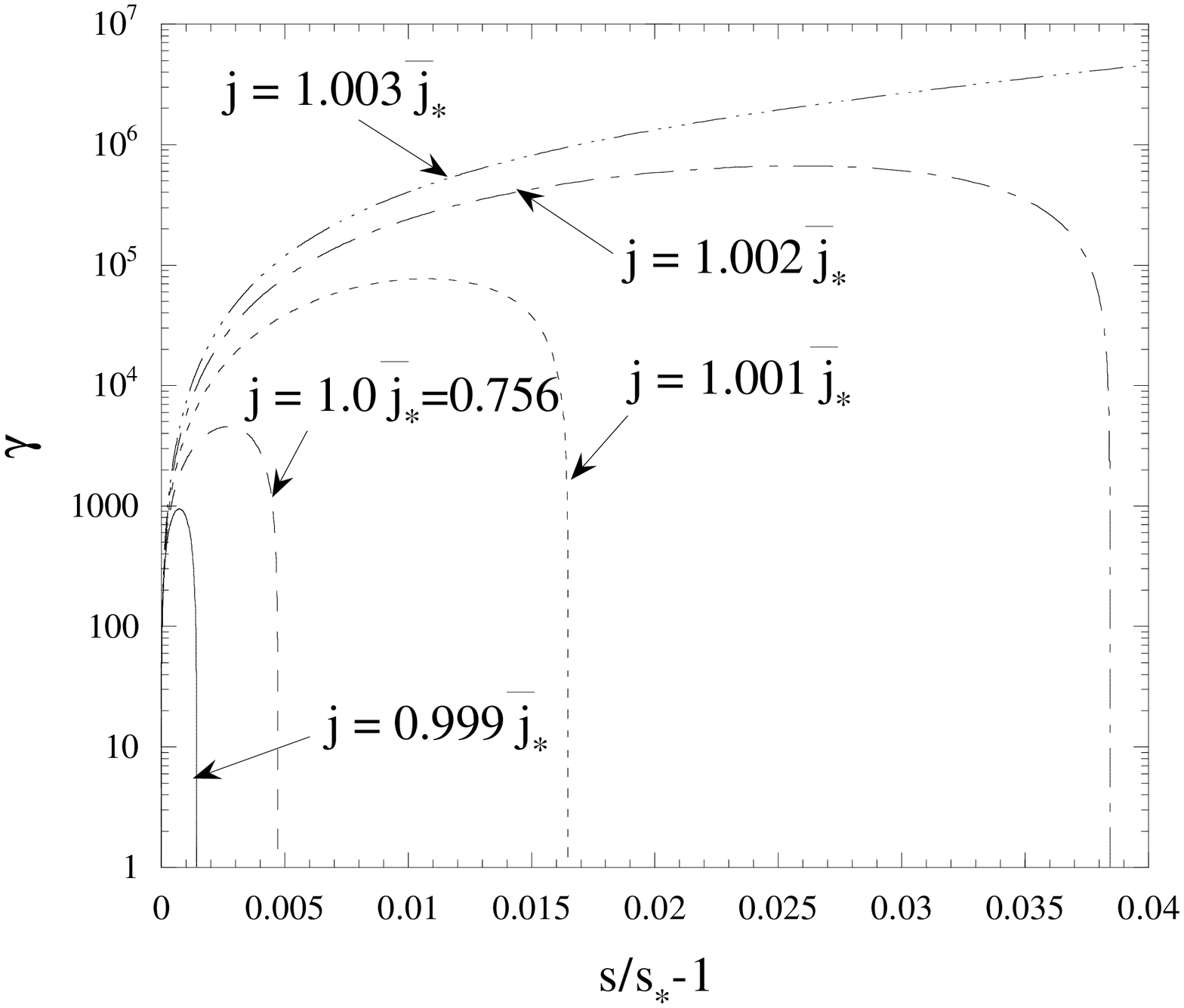}
\caption{Spatial distribution of $\gamma$ along ``away'' field lines. 
We choose $\theta_{\ast}=5^{\circ},~\varphi=180^{\circ}$. 
(a) and (b) represent flat and GR solutions, respectively.
Gravity changes particle dynamics for $j<\bar j_{\ast}$: acceleration after oscillation 
does not occur.}
\end{figure}

\begin{figure}
\figurenum{3}
\epsscale{.9}
\plotone{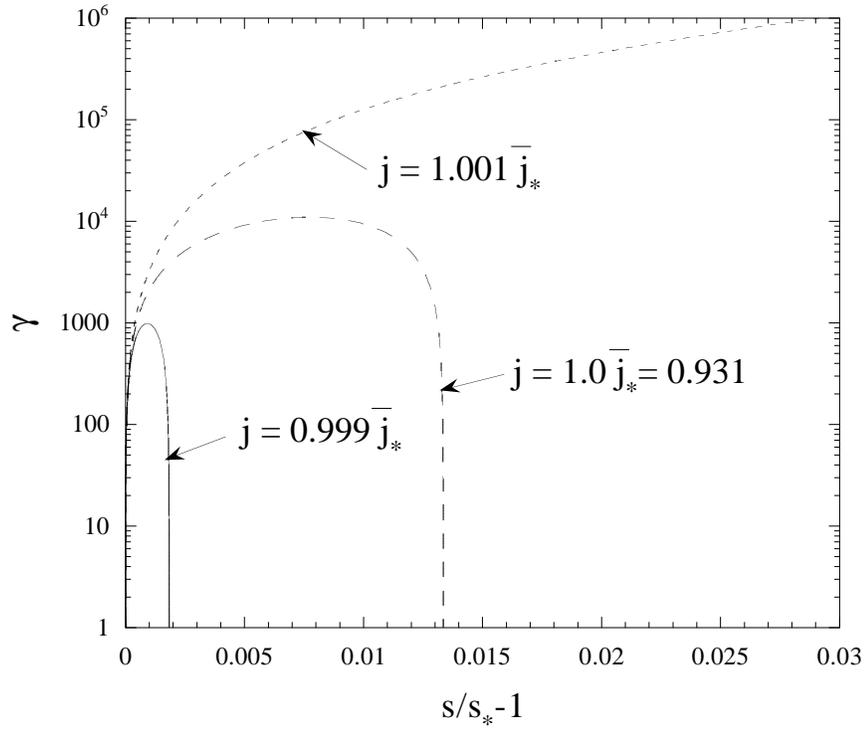}
\epsscale{.9}
\plotone{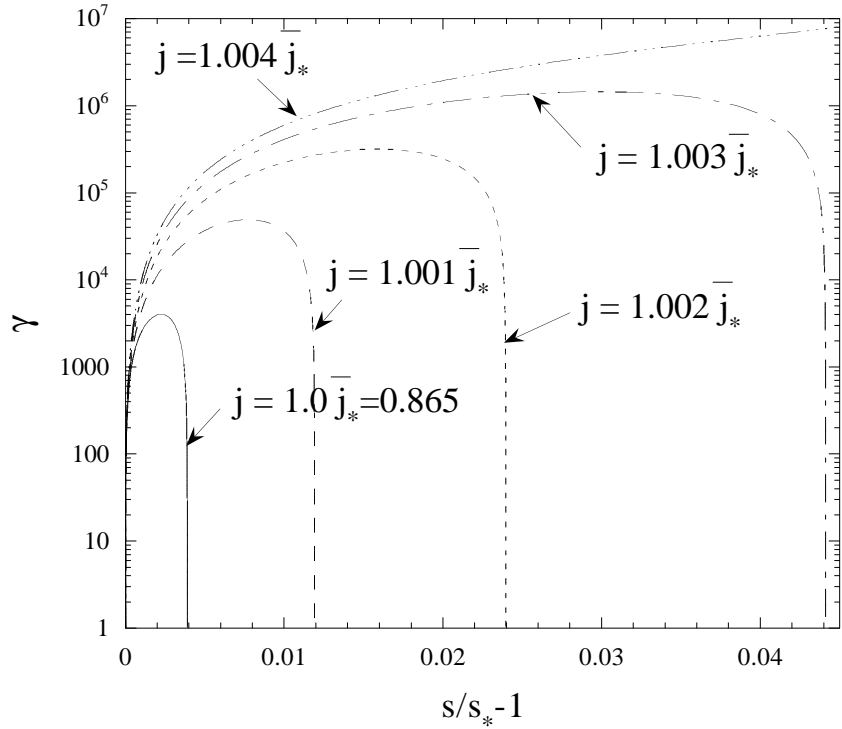}
\caption{Spatial distribution of $\gamma$ along ``toward'' field lines. 
We choose $\theta_{\ast}=5^{\circ},~\varphi=0^{\circ}$.
(a) and (b) represent flat and GR solutions, respectively.}
\end{figure}


\begin{thebibliography}{}
\bibitem[Arons 1997]{Aro}
Arons, J. 1997, in "Neutron Stars and Pulsars" eds. Shibazaki, N.,Kawai, N.,
Shibata, S., \& Kifune, T. (Universal Academy Press, Inc.), 339
\bibitem[Arons \& Scharlemann 1979]{AS}Arons, J., Scharlemann, E.T. 1979, \apj, 231, 854
\bibitem[Ginzburg \& Ozenoi 1965]{GO}
Ginzburg, V.L., \& Ozenoi, L.M. 1965, Sov. Phys. JETP, 20, 689
\bibitem[Goldreich \& Julian 1969]{GJ}
Goldreich, P., \& Julian, W.H. 1969, \apj, 157, 869
\bibitem[Fawley et al. 1977]{FAS}
Fawley, W.M., Arons, J., Scharlemann, E.T. 1977, \apj, 217, 227
Fawley, W.M., Arons, J., Scharlemann, E.T. 1977, \apj, 217, 227
\bibitem[Harding \& Muslimov 2001]{HM}
 Harding, A.K., \& Muslimov, A.G. 2001, \apj, 556, 987
\bibitem[Konno \& Kojima 2000]{KK}
Konno, K., \& Kojima, Y. 2000, Prog. Theor. Phys., 104, 1117
\bibitem[Landau \& Lifshitz 1975]{LL}
Landau, L.D., \& Lifshitz, E.M. 1975, ``The Classical Theory of Fields'' (Pergamon)
\bibitem[Melrose 2000]{Mel}
Melrose, D.B. 2000, in ASP Conf. Ser. 202, ``Pulsar Astronomy: 2000 and beyond", 
eds. M. Kramer, N. Wex, \& R. Wielebinski, 721
\bibitem[Mestel 1981]{Mes81}
Mestel, L. 1981, in IAU Symposium 95, eds. W. Sieber \& R. Wielebinski (Dordrecht, Reidel), 9
\bibitem[Mestel 1996]{Mes96}
Mestel, L. 1996, in ASP Conf. Ser. 105, "Pulsars: Problems and Progress", 
eds. S. Jonston, M.A. Walker, \& M. Bailes (San Francisco: ASP), 417
\bibitem[Muslimov \& Tsygan 1992]{MT}Muslimov, A.G., \& Tsygan, A.I. 1992, MNRAS, 255, 61
\bibitem[Scharlemann, Arons \& Fawley 1978]{SAF}
Scharlemann, E.T., Arons, J., Fawley, W.M. 1978, \apj, 378, 239
\bibitem[Shibata 1991]{Shi91}Shibata, S. 1991, \apj, 276, 537
\bibitem[Shibata 1995]{Shi95}Shibata, S. 1995, MNRAS, 276, 537
\bibitem[Shibata 1997]{Shi97}Shibata, S. 1997, MNRAS, 287, 262
\bibitem[Thorne \& Macdonald 1982]{TM}
Thorne, K.S., \& Macdonald, D. 1982, MNRAS, 198, 339
\bibitem[Weinberg 1972]{Wei}
Weinberg, S. 1972, ``Gravitation and Cosmology'' (John Wiley \& Sons)
\end{thebibliography}
\end{document}